\documentclass[final,journal,a4paper,11pt,twocolumn]{IEEEtran}%
\pdfoutput=1
\usepackage{graphicx}
\usepackage{amsmath}
\usepackage{amsfonts}
\usepackage{amssymb}
\usepackage{balance}
\begin{document}

\title{Adaptive Variable Step Algorithm for Missing Samples Recovery in Sparse Signals
}
\author{Ljubi\v sa Stankovi\' c, Milo\v s Dakovi\' c, Stefan Vujovi\' c
\thanks{Address: University of Montenegro, 81000 Podgorica, Montenegro}
\thanks{E-mail: ljubisa@ac.me, milos@ac.me, stefanv@ac.me}
}
\maketitle
\markboth{Submitted to IET Signal Processing, September 2013}{Submitted to IET Signal Processing, September 2013}

\begin{abstract}
Recovery of arbitrarily positioned samples that are missing in sparse signals
recently attracted significant research interest. Sparse signals with heavily
corrupted arbitrary positioned samples could be analyzed in the same way as
compressive sensed signals by omitting the corrupted samples and considering
them as unavailable during the recovery process. The reconstruction of missing
samples is done by using one of the well known reconstruction algorithms. In
this paper we will propose a very simple and efficient adaptive variable step 
algorithm, applied directly to the concentration measures, without
reformulating the reconstruction problem within the standard linear
programming form. Direct application of the gradient approach to the
nondifferentiable forms of measures lead us to introduce a variable step size
algorithm. A criterion for changing adaptive algorithm parameters is presented.
The results are illustrated on the examples with sparse signals, including
approximately sparse signals and noisy sparse signals.


\end{abstract}

\section{Introduction}

In many signal processing applications a signal that spans over the whole time
domain is located within much smaller regions in a transformation domain. If
we consider a discrete time-limited signal, it could contain much smaller
number of nonzero samples (coefficients) in an arbitrary transformation domain
(Fourier domain, Discrete cosine domain, Discrete Wavelet domain,...). It is
said that this signal is sparse in this transformation domain. Remaining
transformation coefficients of the signal are assumed to be equal to zero or
could be approximated by zero without making significant error. If this
condition is satisfied we can reconstruct the signal without using the whole
data set required by the Shannon-Nyquist sampling theorem. Processing of
sparse signal with a large number of missing/unavailable samples attracted
significant interest in recent years. This research area interacts with many
other research areas like signal processing, statistics, machine learning,
coding. Compressive sensing/sampling (CS) is a field which refers to a sparse
approximation \cite{donoho2006,candes2006}. The crucial parameter in the
approximation is the number of available samples/measurements used in the
reconstruction. It is directly related to the number of non-zero sparse
coefficients \cite{donoho2006,candes2006,mallat1999}.

Signal samples may be missing due to their physical or measurements
unavailability. Also, if some arbitrary positioned samples of signal are so
heavily corrupted by disturbance, it was shown that it is better to omit them
in the analysis or processing (by L-estimation, for example \cite{22, 7a,
knjiga}). Both of these situations corresponds to the CS approximation
problems if the signal to be analyzed is sparse and could be considered within
the framework of missing samples. Under the conditions defined within the CS
framework, the processing of a signal could be performed with the remaining samples almost
as in the case if all missing/unavailable samples were available.

Several approaches to reconstruct these kind of signals are introduced
 \cite{Gradient1}-\cite{figueiredo2007}. One group
of them is based on the gradient \cite{figueiredo2007} and the other is based
on the matching pursuit approaches \cite{mallat1993}. 
A common approach to this problem is based on redefining it within the linear programming (LP) as the bound constrained quadratic program (BCQP).
A measure of
signal sparsity is used as a minimization function in the sparse signal
reconstruction. This measure is related to the number of nonzero transformation
coefficients. This kind of measure was also used, especially in
time-frequency analysis, for measuring the concentration of a signal
representation. Since, the sparsity of a signal in a transformation domain is
related to the number of nonzero samples, a natural mathematical form to
measure the number of nonzero (significant) samples in a signal transform is
the norm-zero ($l_{0}$ norm). This norm is sum of the signal transformation
absolute values raised to a zeroth power. Since this power produces value one
for any nonzero transformation coefficient, the norm just counts the number of
nonzero coefficients. However, this norm is very sensitive to any kind of
disturbance that can make zero transformation values to be small but different
from zero. Thus more robust norms are used. The norm that may be used for
measuring the transformation concentration, being also less sensitive to
disturbances, is the norm-one ($l_{1}$ norm). Since the norm-one is not
differentiable around the optimal point, the CS algorithms reformulate problem
under the norm-two ($l_{2}$ norm) conditions before the optimization is done.

In this paper we will present a gradient based algorithm for reconstruction of
sparse signals. Presented algorithm uses an arbitrarily concentration measure
in a direct way, without redefining the problem to a quadratic form and using
liner programming tools. Since the signal reconstruction is required in the
time domain, the proposed algorithm performs a search over the time domain
coefficients. The presented algorithm can reconstruct a large number of
missing samples in computationally efficient way. The proposed method belongs
to the class of gradient based CS algorithms \cite{Gradient2}. However, common
adaptive signal processing and the CS algorithms avoid direct use of the
measure based on the norm-one (or similar norms between norm-zero and
norm-one) since it is not differentiable. The value of gradient cannot be used
as an indicator of the proximity of iteration values to the algorithm
solution. When the iterations are close to the optimal point, gradient value
remains the same in norm-one and oscillate around the true value. Taking
sufficiently small step over the whole range would not be a solution, due to
extremely large number of iterations over a very large set of variable. Here,
the adaptive gradient based approach is directly applied to an appropriately
chosen concentration measure. Since for commonly used norm-one (or any other
norm between norm-zero and norm-one)  based concentration measure the
derivatives are not continuous functions around the minimum, a variable and
self-adaptive step in the algorithm is introduced. The presented algorithm,
with this adaptive step, reconstructs a large number of missing samples in a
simple and computationally efficient way with arbitrary (computer defined)
precision of the results. 

The paper is organized as follows. After the introduction, a review and
analysis of concentration measures in the processing of sparse signals is
done. A gradient based algorithm, with its modifications is presented and
illustrated. The presented algorithm efficiency is demonstrated on several
examples with large number of missing samples, including the samples missing in
blocks and the noisy signals. The basic idea for this algorithm was presented 
in \cite{trst2013}.

\section{Measures and Direct reconstruction}

Concentration measures of signal transforms were intensively studied and used
in the area of time-frequency signal analysis and processing. They are used to
find an optimal, best concentrated time-frequency representation of a signal.
The most common and the oldest measure introduced to measure concentration of
time-frequency representations is defined 
by  Jones, Parks, Baraniuk, Flandrin, Williams, \textit{et al.} Concentration
of a signal transform $X(k)$ is measured by%
\begin{equation}
\mathcal{M}^{(4/2)}[X(k)]=\frac{\sum\nolimits_{k}\left\vert X(k)\right\vert
^{4}}{\left(  \sum\nolimits_{k}\left\vert X(k)\right\vert ^{2}\right)  ^{2}%
}.\label{OJP}%
\end{equation}
In general, it has been shown that any other ratio of norms $l_{p}%
=\sum\nolimits_{k}\left\vert X(k)\right\vert ^{p}$ and $l_{q}=\sum
\nolimits_{k}\left\vert X(k)\right\vert ^{q}$, $p>q>1$, can also be used for
measuring the concentration. This kind of concentration measures were inspired
by the kurtosis as a measure of distribution peakedness. Similar forms are
obtained by using the R\'{e}nyi measures.

 Another
direction to measure time-frequency representation concentration comes from a
classical definition of the time-limited signal duration, rather than
measuring signal peakedness. It was used in time-frequency analysis in
\cite{Ljubisa}. If a signal $x(n)$ is time-limited, $x(n)\neq0$ only for $n\in\lbrack n_{1},n_{2}-1]$,
then the duration of $x(n)$ is $d=n_{2}-n_{1}$. It can be written as
\begin{equation}
d=\lim_{p\rightarrow\infty}\sum\nolimits_{n}\left\vert x(n)\right\vert
^{1/p}=\left\Vert x(n)\right\Vert _{0},\label{trajanje}%
\end{equation}
where $\left\Vert x(n)\right\Vert _{0}$ denotes the norm-zero $l_{0}$ of
signal. In reality, there is no sharp edge between $x(n)\neq0$ and $x(n)=0$,
so the value of $d$ in (\ref{trajanje}) could, for very large $p$ (close to
norm-zero), be sensitive to small values of $\left\vert x(n)\right\vert $. The
robustness may be achieved by using lower-order forms, with $1\leq p<\infty$
(norms from $l_{1}$ to $l_{0}$).

Therefore, the concentration of a signal transform $X(k)=T[x(n)]$ can be
measured with the function of the form
\begin{equation}
\mathcal{M}_{p}[T[x(n)]]=\frac{1}{N}\sum\nolimits_{k}\left\vert
X(k)\right\vert ^{1/p},\label{povd}%
\end{equation}
with  $1\leq p<\infty$, where $N$ is total number of samples in signal
transform $X(k)$. A lower value of (\ref{povd}) indicates better concentrated
distribution. For $p=1$, it is the norm-one form
\[
\mathcal{M}_{1}[T[x(n)]]=\frac{1}{N}\sum\nolimits_{k}\left\vert
X(k)\right\vert =\frac{1}{N}\left\Vert X(k)\right\Vert _{1}.
\]
Minimization of the norm-one of the short-time Fourier transform (norm 1/2 of the spectrogram) is used in \cite{Ljubisa}  to optimize the window width and to produce the best concentrated signal representation.  The norm-one is also the most commonly used in the CS algorithms for measuring signal sparsity/concentration \cite{donoho2006,candes2006,5,6}. Here
 we will illustrate the influence of measure parameter $p$
on the results, including explanation why the norms greater than one ($p<1,$
including $l_{2}$ case) cannot be used for measuring concentration. These
norms could be used in the ratio forms (\ref{OJP}) only, but not as the
stand-alone transformation measures, like in the case of $1\leq p<\infty$.

The simplest reconstruction algorithm will be based on a direct search over
all unavailable/missing samples values, by minimizing the concentration
measure. If we consider a complete set of signal samples 
$
\left\{  x(1),x(2),...,x(N-1)\right\}
$
and $M$ samples $x(m_{1})$, $x(m_{2})$,...,$x(m_{M})$ are missing then the
simplest algorithm will be to search over all possible values of missing
samples and find solution that minimizes the concentration measure%
\[
\underset{x(m_{1}),x(m_{2}),...,x(m_{M})}{\min}\left\{  \mathcal{M}%
_{p}[T[x(n)]]\right\}  .
\]
From the remaining samples we can estimate the range limits for the missing
samples, $|x(m_{k})|\leq A$. In the direct search approach we can vary each
missing sample value from $-A$ to $A$ with a step $2A/(L-1)$ where $L$ is the
number of considered values within the selected range. It is obvious that the
reconstruction error is limited by the step $2A/(L-1)$ used in the direct
search. Number of analyzed values for $M$ coefficients is $L^{M}$. Obviously,
the direct search can be used only for a small number of missing samples, since
for any reasonable accuracy value of $L$ is large.

One possible approach to reduce the number of operations in the direct search
is to use a large step (small $L$) in the first (rough) estimation, then to
reduce the step around the rough estimate of unavailable/missing values
$x(m_{1})$, $x(m_{2})$,..., $x(m_{M})$. This can be repeated several times,
until the desired accuracy is achieved. For example, for $A=1$ the accuracy of
$0.001$ is achieved in one iteration if $L=2001$. With, for example, 7 missing samples that
would mean unacceptable number of $2001^{7}\sim10^{23}$ measure
calculations. However, if the first search is done with, for example $L=5$,
the rough optimal values are found, and the procedure is repeated with $L=5$
values within the range determined by the rough optimal in the first step.
Repeating the same procedure six more times, the accuracy better than $0.001$
is reached with $7\times5^{7}\sim10^{5}$ measure calculations. In this
way, we have been able to analyze (on an ordinary PC, within a reasonable
calculation time), signals with up to $10$ missing samples.

Although, computationally not efficient, the direct search method is very important
and helpful in the analysis of various concentration measures with different
$p$, since all more advanced and efficient methods from literature produce
results with nice values of $p$ only (for example, $p=1$, $p=1/2,$ or $p=2$).
The direct method can be used with any $p$. Also, the probability that we find
and stay in a local minimum is lower in the direct method than when using
other algorithms. Thus, we will use the direct search to illustrate how the
solution depends on the chose norm (concentration measure form).

\begin{figure}[tp]
\centering \includegraphics{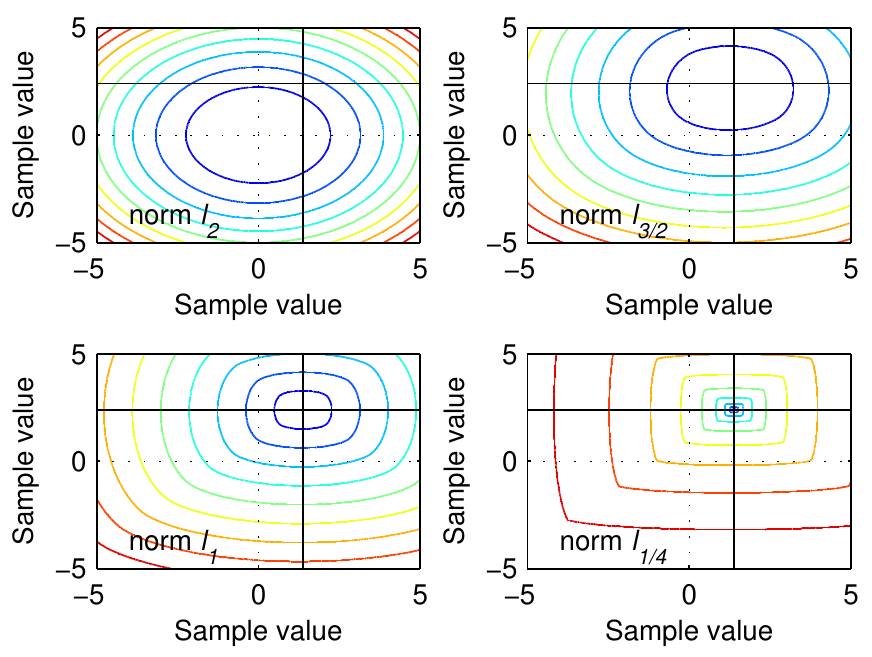} \caption{Measure as a function of
two missing sample values for various norms. True values of missing samples are
presented with black lines}%
\label{mjereXXX_2D}%
\end{figure}

\begin{figure}[tp]
\centering \includegraphics{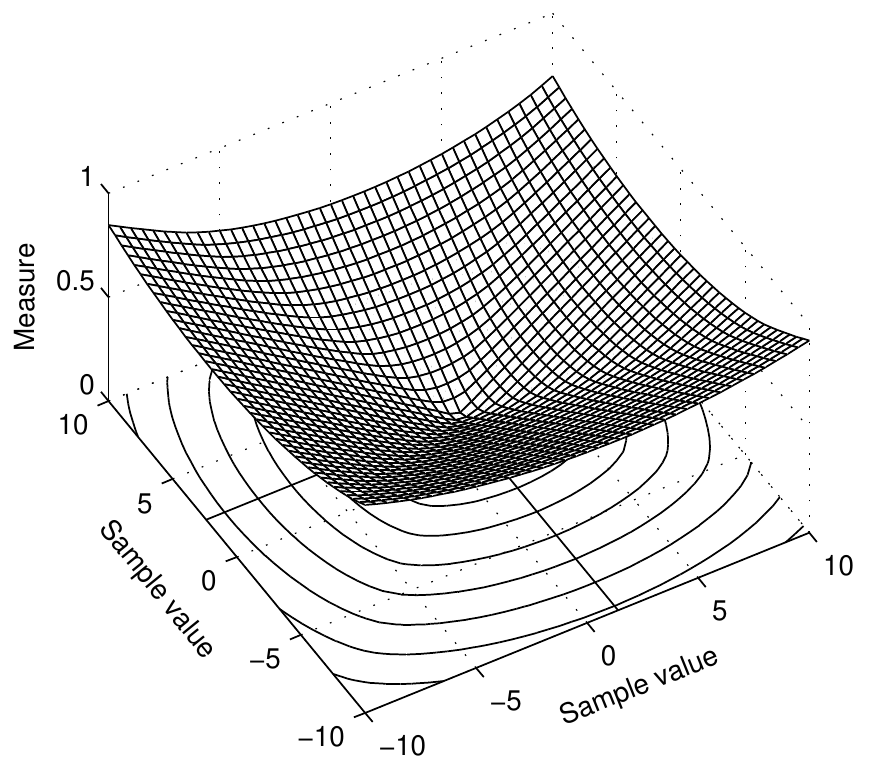} \caption{Measure for $p=1$ (norm $l_1$) as a
function of the missing sample values}%
\label{mjereXXX_2D_Mesh}%
\end{figure}

\textit{Example}: Consider a discrete signal
\begin{equation}
x(n)=2.5\sin(20\pi n/N)\label{SigEx1}%
\end{equation}
for $n=0,1,\ldots,N-1$, and $N=256$ is number of signal samples. The case of
two missing samples is presented first, as the one appropriate for graphical
illustration. Direct search is performed over range $[-5,5]$ with step $0.01$.
The measure (\ref{povd}) is calculated for various values of parameter $p$.
Results are shown in Fig.~\ref{mjereXXX_2D}. The measure minimum is located on
the true sample values for $p\geq1$ (norms $l_{1}$ and lower). The measure
minimum for $p<1$ (including norm-two, $l_{2}$, for $p=1/2$) is not located at
the true signal values. Important case with two missing samples and $p=1$ is
presented in Fig.~\ref{mjereXXX_2D_Mesh}. 

In order to illustrate the measure influence on the mean absolute error (MAE)
the direct search is also performed on the signal
\begin{equation}
x(n)= 
3\sin(10\pi n/N)+2\cos(30 \pi n/N).
\label{SigExample2}%
\end{equation}
Signal is composed of $N=256$ samples while the cases with 4 and 7 missing
samples are analyzed. The results with $10$ and $15$ iterations (to reduce the
step size) are presented in Fig.~\ref{MAE_DIREC_4}. We can see that:

\begin{figure*}[tp]
\centering \includegraphics{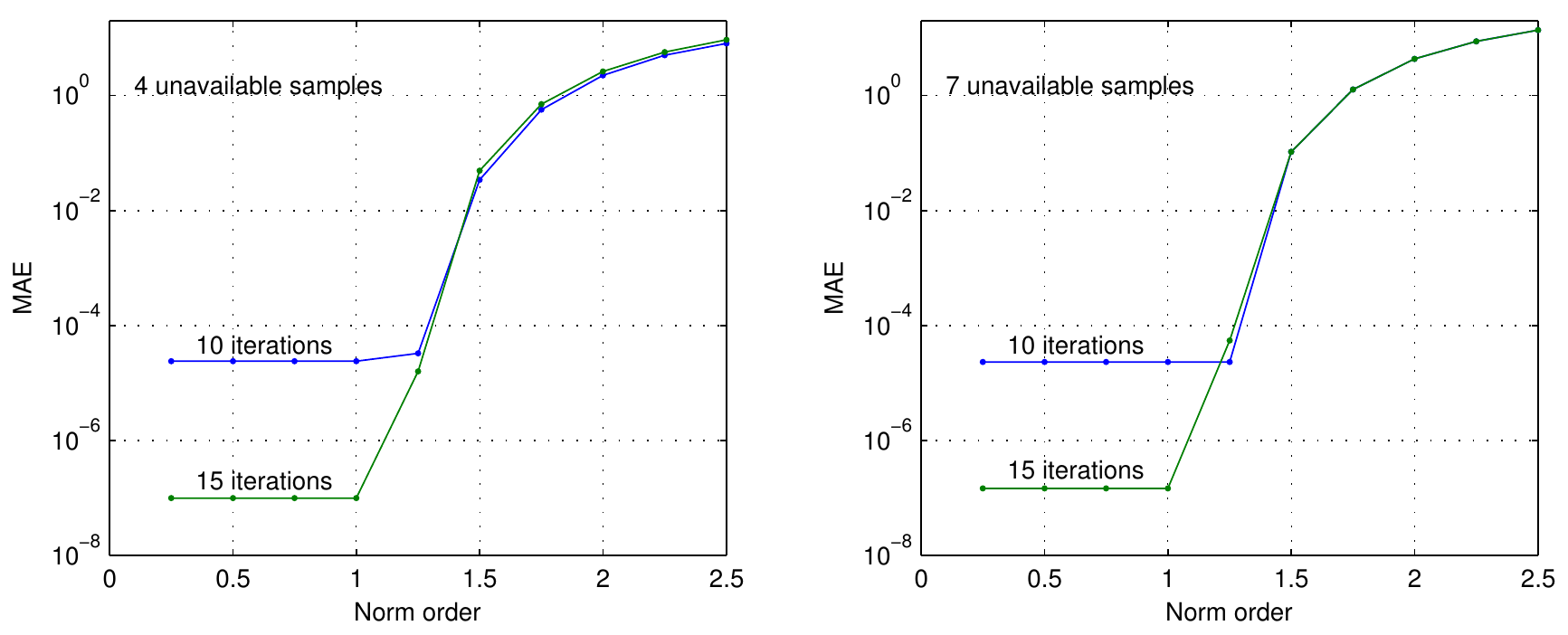} \caption{Mean absolute
error (MAE) in the coefficients estimation as a function of the norm $l$ order
 for $4$ missing samples (left) and $7$ missing samples (right). The MAE 
 is normalized with the
number of missing samples.}%
\label{MAE_DIREC_4}%
\end{figure*}

1) $p\geq1$ (norms $l_{q}$ with $q=1/p\leq1$, including $l_{1}$) produces
accurate results with the MAE depending on the direct search step only. The
MAE can be further reduced to the computer precision by reducing the step size
with more iterations.

2) For $p<1$ (norms $l_{q}$ with $q>1$, including $l_{2}$) the bias dominate
over the number of iterations, so the results are almost independent from
number of iterations.

Almost the same results are obtained for 4 and 7 missing samples cases.

\textit{Minimization Using }$l_{2}$\textit{ Norm: }For $p=0.5$ this measure is
equivalent with the well-known $l_{2}$ norm used in definitions of standard
signal transforms \cite{knjiga}. For the norm-two $(l_{2}$ norm with $p=1/2)$
the MAE is of the signal samples order, as shown in Fig.~\ref{MAE_DIREC_4}.
\textbf{The measure with }$l_{2}$\textbf{ norm has a minimum when the missing
signal samples values are set to zero.} 

This result was expected and can be proven for any number of missing samples for signals and its transforms satisfying the Parseval's theorem.
The Parserval's theorem states that the energy of a signal in the time domain
is the same as the energy of the Fourier transform in the frequency domain. We
know that signal has the lowest energy when its missing samples are
zero-valued. Adding any other value of missing samples than zero will increase
the energy. The same holds in the frequency domain since the energy in
the frequency domain equals the energy in the time domain, for this norm. The
minimization solution with the $l_{2}$ norm is therefore trivial. With this
norm, we attempt to minimize
\[
\left\Vert \mathbf{X}\right\Vert _{l_{2}}=\sum_{k=0}^{N-1}\left\vert
X(k)\right\vert ^{2}.
\]
According to Parseval's theorem we have $\left\Vert \mathbf{X}\right\Vert
_{l_{2}}=N\sum_{n=0}^{N-1}\left\vert x(n)\right\vert ^{2}$. Since any value
other than $x(n)=0$ for the unavailable/missing signal samples, would
increase $\left\Vert \mathbf{X}\right\Vert _{2}$, then the solution for the
non-available sample values, with respect to the $l_{2}$ norm, are all zero
values. This was the reason why this norm was not used as a concentration
measure. This is also the reason why this norm can not be used in the CS based
algorithms for the missing samples recovery.

\section{Adaptive Gradient Based Algorithm}

Due to a high computational complexity the direct search could be used only if
the number of missing samples $M$ is small enough. It is the reason why many
other, more sophisticated, CS algorithms have been proposed. Here, we will
present one very simple and efficient algorithm, based on the direct use of
the concentration measure gradient. This algorithm is inspired by the adaptive
signal processing methods with a variable step size. This algorithm is the
form of gradient descent algorithm where missing samples are estimated as the
ones producing best concentration measure of the signal transformation in the
sparse domain.

The norms that produce unbiased missing samples values (like for example norms
with $p\geq1$) are not differentiable around the optimal point. It means that
the gradient method, if directly applied to the measure based on, for example,
the $l_{1}$ norm (or any other norm with $p\geq1$), will have a problem when approaching
the optimal point. Since the gradient amplitude in the vicinity of
the optimal point is almost constant for $p=1$ (with a changing sign), the
algorithm will not improve the accuracy to a level lower than the accuracy
defined by the step in the gradient algorithm. This is the reason why this
approach has not been used and why appropriate reformulation of this problem is
done in literature. These reformulations are done within the linear
programming by using the well known and available norm-two based solutions.
Here, we will not try to reformulate the problem based on the $l_{1}$ norm (or
any other norm with $p\geq1$) within the linear programming $l_{2}$ framework,
but to use the gradient based adaptive algorithm, with the step being
appropriately adjusted (in a simple way) around the optimal point. The
algorithm presented here will be a simple (and efficient) direct application
of the gradient based adaptive approach to the measures based on norms that
are not differentiable around the optimal point, like the $l_{1}$ norm.

As we can see from Fig.~\ref{mjereXXX_2D_Mesh} measure with $p=1$ is
differentiable and convex everywhere except around the point of minimum (the
optimization solution point). Therefore any algorithm applied directly to the
measure based on $p=1$ will oscillate around the solution with an amplitude
defined by the step and measure form (this will be illustrated within
examples). If we take a very small step for each of a large number of missing
samples, it will result in an unacceptable and large number of iterations.
Thus, when the steady oscillatory state (steady state in mean absolute error)
is detected we should reduce the algorithm step, as we presented in the direct
search. In this way, the results with a high accuracy, within an acceptable
number of iterations, are achieved with a variable self-adaptive step. This
simple method is able to produce the results with an error of the computer
precision level. Finally, in addition to the step variation, this kind of
algorithms enables that the parameter $p$ (the norm form itself) is changed to
improve the initial convergence of the algorithm.

\subsection{Algorithm}

Consider a discrete signal $x(n)$ with some samples that are not available.
Assume that signal is sparse in a transformation domain $T[x(n)]$. The
algorithm for missing samples reconstruction is implemented as follows:

\noindent\textbf{Step 0:} Form the initial signal $y^{(0)}(n)$, where $(0)$
means that it is first iteration of algorithm, as:
\[
y^{(0)}(n)=\left\{
\begin{array}
[c]{ll}%
x(n) & \text{ for available samples}\\
0 & \text{ for missing samples}%
\end{array}
\right.
\]

\noindent\textbf{Step 1:} For each missing sample at $n_{i}$ we form two
signals $y_{1}(n)$ and $y_{2}(n)$ in each next iteration as
\[
y_{1}^{(k)}(n)=\left\{
\begin{array}
[c]{ll}%
y^{(k)}(n)+\Delta & \quad\text{for $n=n_{i}$}\\
y^{(k)}(n) & \quad\text{for $n\neq n_{i}$}%
\end{array}
\right.
\]%
\[
y_{2}^{(k)}(n)=\left\{
\begin{array}
[c]{ll}%
y^{(k)}(n)-\Delta & \quad\text{for $n=n_{i}$}\\
y^{(k)}(n) & \quad\text{for $n\neq n_{i}$}%
\end{array}
\right.
\]
where \textit{k} is the iteration number. Constant $\Delta$ is used to
determine whether the considered signal sample should be decreased or increased.

\noindent\textbf{Step 2:} Estimate the differential of the signal transform
measure as
\begin{equation}
g(n_{i})=\frac{\mathcal{M}_{p}\left[  T[y_{1}^{(k)}(n)]\right]  -\mathcal{M}%
_{p}\left[  T[y_{2}^{(k)}(n)]\right]  }{2\Delta}, \label{eq:mjera}%
\end{equation}
where $\mathcal{M}_{p}$ is defined by (\ref{povd}). The differential of
measure is proportional to the error ($y^{(k)}(n)-x(n)$).

\noindent\textbf{Step 3:} Form a gradient vector $\mathbf{G}$ with the same
length as signal $x(n)$. At positions of available samples, this vector has
value $G(n)=0$. At the positions of missing samples it has values are
$g(n_{i})$ calculated by (\ref{eq:mjera}).

\noindent\textbf{Step 4:} Correct the values of signal $y(n)$ iteratively by
\[
y^{(k+1)}(n)=y^{(k)}(n)-\mu G(n)
\]
where $\mu$ is a constant that affect performances of algorithm (error and
speed of convergence).

Repeating the presented iterative procedure, the missing values are going to
converge to the true signal values that produce minimal concentration measure
in the transformation domain. The algorithm performance depend on parameters
$\mu$ and $\Delta$. Here we will use measure that are close to norm-one based measure.

\subsection{Varying and Adaptive Step Size}
Since we use a difference of the
measures to estimate the gradient, when we approach to the optimal point, the
gradient with norm $l_{1}$ will be constant and we will not be able to
approach the solution with a precision higher than the step $\mu,$ multiplied
by constant (gradient dependent value). If we try to reduce the oscillations
around the true value by using smaller step from the beginning, then we will
face with an unacceptable number of iterations. However, this problem may be
solved, by reducing the step size, when we approach the stationary
oscillations zone. The best solution is to use an adaptive step size in the
algorithm. A large step size should be used when the concentration measure is
not close to its minimum (in the starting iterations). The step is reduced as
we approach the concentration measure minimum. Next we will present a method
for adaptive parameters adjustment that could be applied to the proposed
algorithm in order to reduce the error and to increase accuracy. 

When the algorithm with constant parameters is close to the optimal point the concentration measure tends to have constant value
in a few consecutive iterations. This behavior will be detected by checking the difference
between two consecutive measure calculations
$
\mathcal{M}_{p}^{(k-1)}-\mathcal{M}_{p}^{(k)}
$, 
where $\mathcal{M}_{p}^{(k)}=\mathcal{M}_{p}\left[  T[y^{(k)}(n)]\right]$ is the sparsity measure of the reconstructed signal in the $k$-th iteration.
 When it is smaller than, for example 
one percent of the highest previously calculated measure difference,
\begin{equation}
\mathcal{M}_{p}^{(k-1)}-\mathcal{M}_{p}^{(k)}\le {P}\max
_{m=1,2,...,k-1}\left\vert \mathcal{M}_{p}^{(m-1)}-\mathcal{M}_{p}%
^{(m)}\right\vert,
\label{krit}
\end{equation}
 where $P=0.01$, the algorithm parameters $\Delta$ and $\mu$ should be reduced, for example 10 times.

\section{Numerical Examples}

\textit{Example 1: }Consider the signal 
\begin{align}
x(n)= &  3\sin(20\pi n/N)+2\cos(60\pi n/N)\nonumber\\
&  +0.5\sin(110\pi n/N).\label{SigExample2b}%
\end{align}
 The total number of signal samples is $N=256$. We assume that $200$ samples are missing or are not available. Two cases will be considered. One when missing samples are randomly positioned and other when samples are missing in randomly positioned blocks. 
We know their positions, as well as that the signal is sparse in the Fourier
domain. 
Here we will perform the signal reconstruction with  constant algorithm parameters $\Delta=2$, $\mu=3$, and $p=1$.

\begin{figure*}[tp]
\centering  \includegraphics{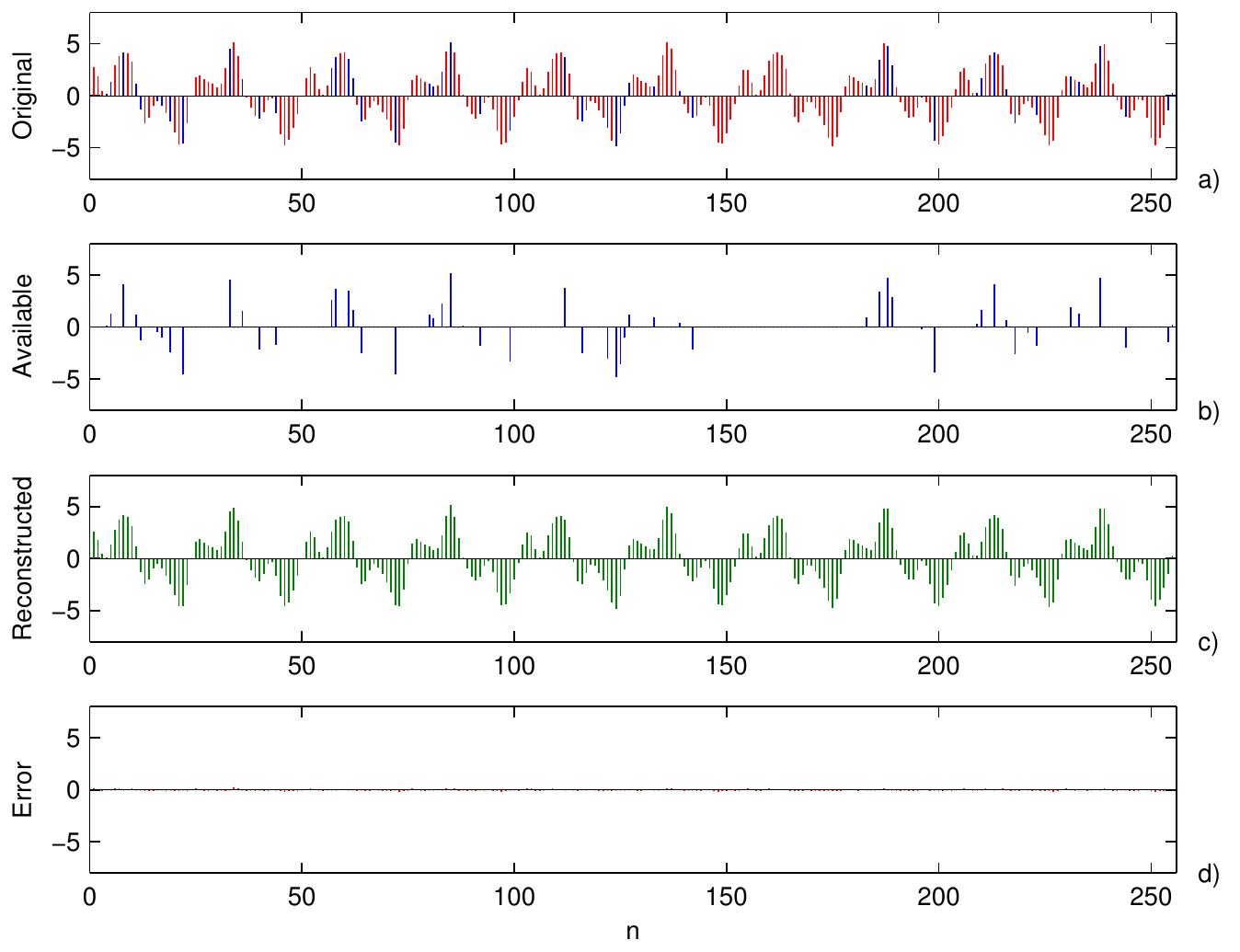} \caption{Reconstruction
example for signal with 200 missing samples at random positions. (a)
original signal; (b) signal with missing samples set to 0 and used as an input to the
reconstruction algorithm; (c) reconstructed signal;  (d) reconstruction error.}%
\label{rekonstrukcija}%
\end{figure*}

\begin{figure*}[tp]
\centering \includegraphics{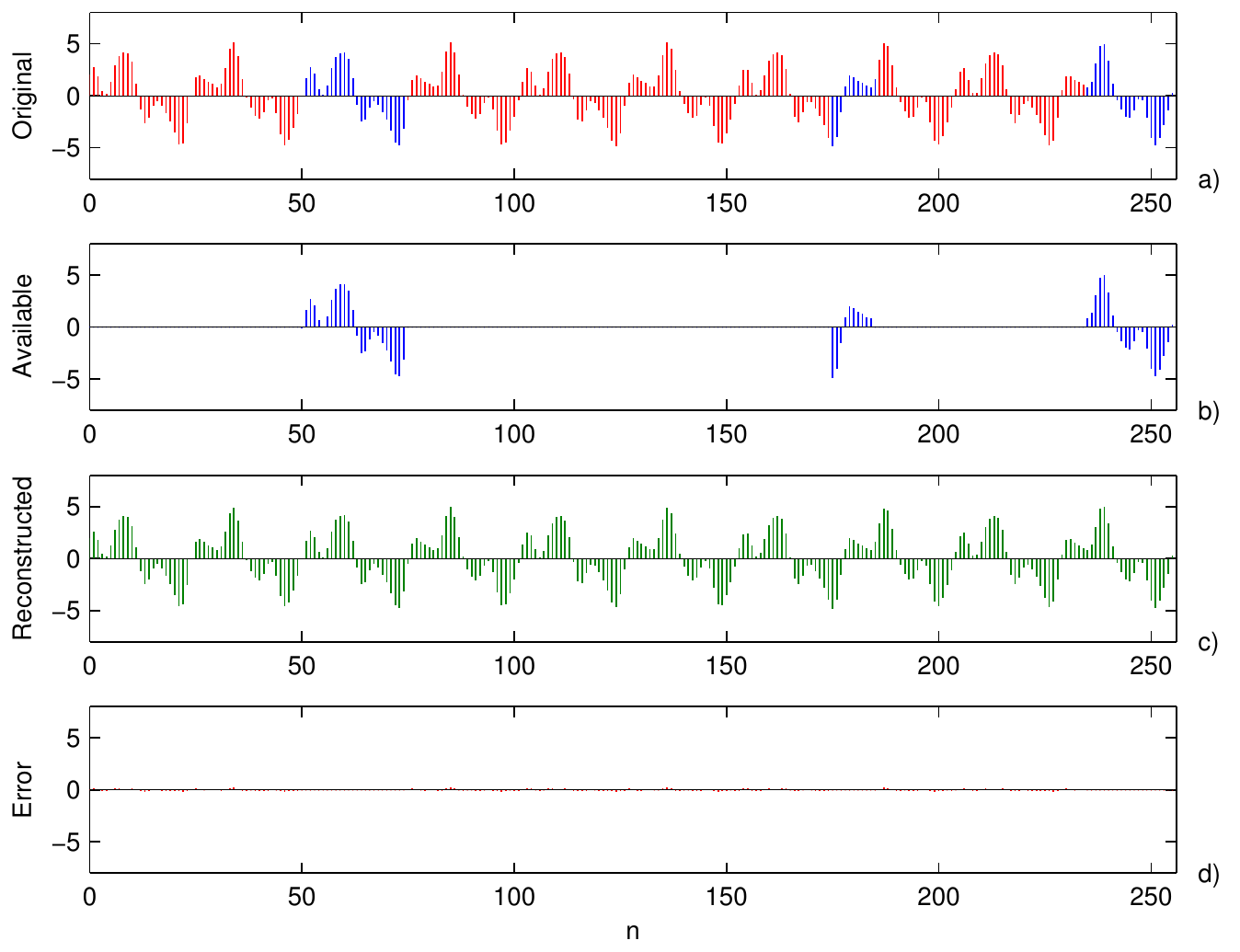}\caption{Reconstruction
example for a signal with 200 missing samples grouped into three blocks.
(a) original signal; (b) signal with missing samples set to 0 and used as an input to the
reconstruction algorithm; (c) reconstructed signal; (d) reconstruction error.}%
\label{rekonstrukcija200}%
\end{figure*}

\begin{figure*}[tp]
\centering \includegraphics{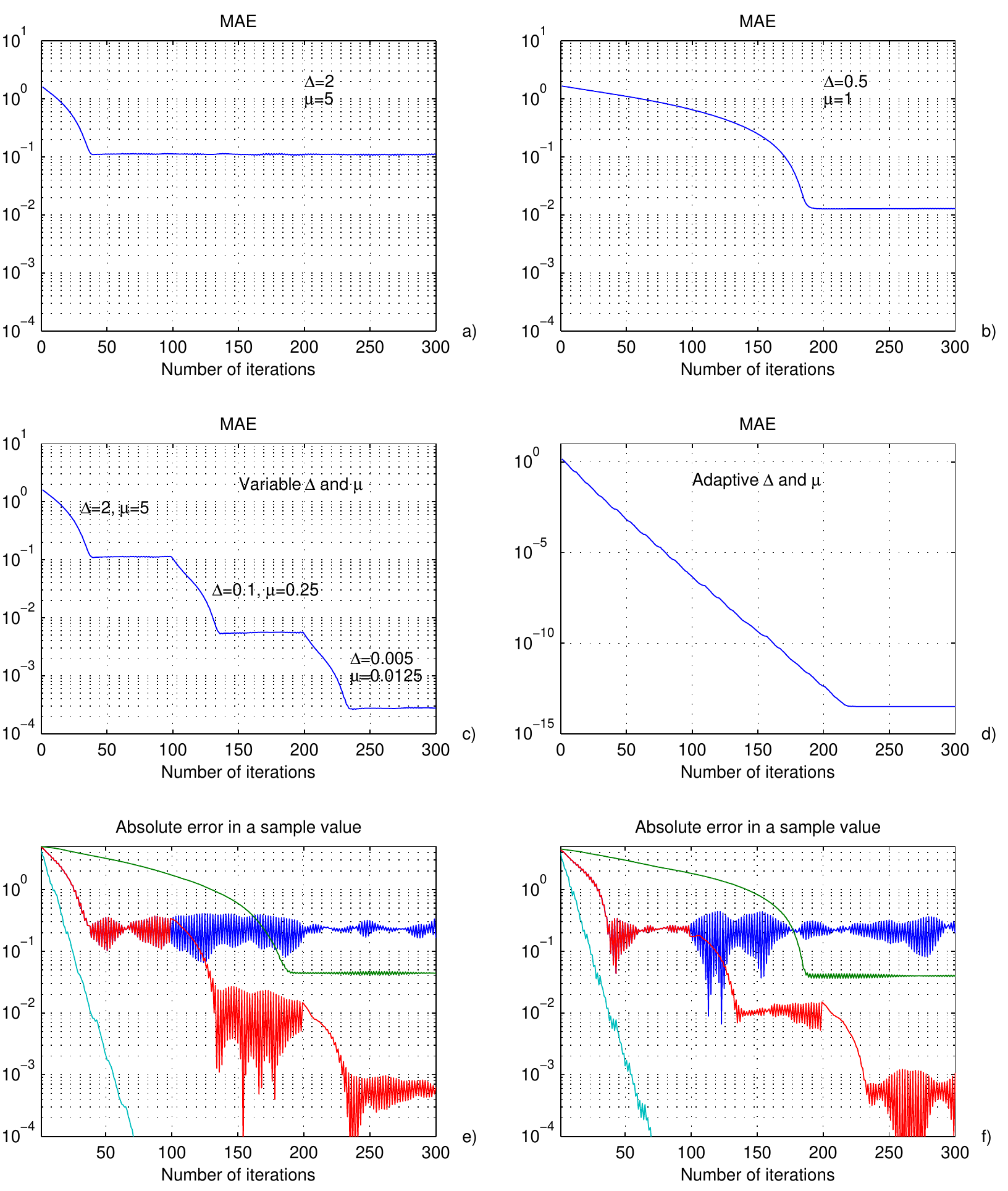} \caption{Mean absolute error (MAE) for constant algorithm parameters $\Delta$ and $\mu$ (a)
and (b); variable parameters and adaptive parameters (c) and (d); absolute
errors for two randomly chosen missing signal samples (e) and (f) for constant
algorithm parameters (green and blue line), variable (red line), and adaptive
parameters (bright blue). }%
\label{MAE_DeltaMiPPP_COEF}%
\end{figure*}

The reconstruction
results are shown in Fig.~\ref{rekonstrukcija} and Fig.~\ref{rekonstrukcija200}. Since the constant algorithm parameters are used the achieved error is small but still notable, Fig.~\ref{rekonstrukcija}(d), Fig.~\ref{rekonstrukcija200}(d). 
The residual error value is determined by the algorithm parameters. Next we will analyze influence of parameters $\Delta$, $\mu$, $p$ on the number of iterations and mean absolute error (MAE). The MAE calculated
as
\[
\mathrm{MAE}(k)=\frac{1}{N}\sum_{n}|x(n)-y^{(k)}(n)|
\]
is shown in Fig.~\ref{MAE_DeltaMiPPP_COEF} for various algorithm setups. It
can be concluded that for constant algorithm parameters $\mathrm{MAE}$ cannot
be improved by increasing number of iterations below some limit. Smaller
values of $\Delta$ and $\mu$ produce lower $\mathrm{MAE}$ but with an
increased number of iterations, as presented in Fig.~\ref{MAE_DeltaMiPPP_COEF}%
(a) and (b). The results obtained for varying $\Delta$ and $\mu$ are presented
in Fig.~\ref{MAE_DeltaMiPPP_COEF}(c). Here, the parameters are changed at
iterations $k=100$ and $k=200$. We can see that with the same number of
iterations a smaller $\mathrm{MAE}$ is achieved. Therefore, the parameters
$\Delta$ and $\mu$ should be adaptive, resulting in $\mathrm{MAE}$ presented
in Fig.~\ref{MAE_DeltaMiPPP_COEF}(d). Here we detect that after some number of
iterations the gradient algorithm does not further improve sparsity of the
reconstructed signal and then we use smaller values of $\Delta$ and $\mu$ for
next iterations, as described in the previous section. In the
Fig.~\ref{MAE_DeltaMiPPP_COEF}(e) and (f) the absolute errors in two signal
samples, during iteration process, are shown for all previous cases of the
algorithm setup. It can be seen that this absolute errors behave in a similar
manner as the MAE in the above subplots, with difference that they oscillate
around the steady value, due to nondifferentiable measure around the solution.
It is expected, as previously explained.

\begin{figure*}[tp]
\centering\includegraphics{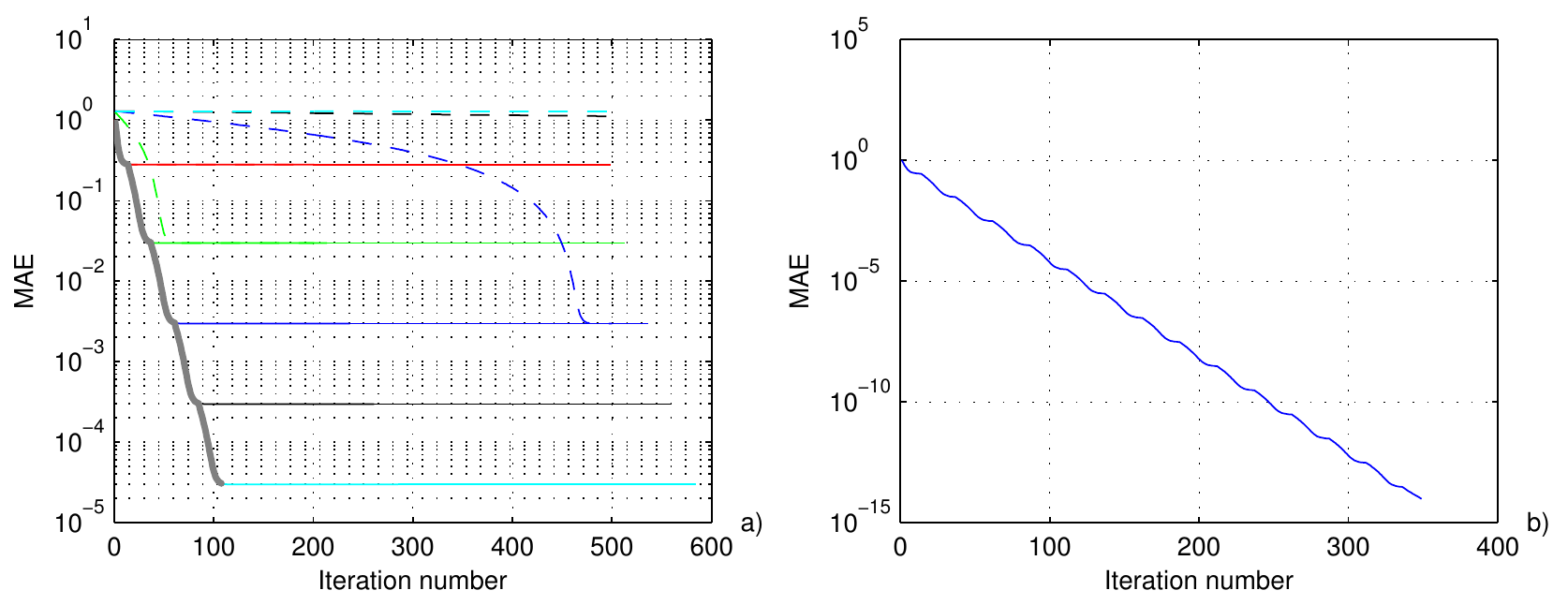}\caption{MAE for adaptive  parameters using the measure-based criterion.
Each color on graphics a) presents one set of parameters $\Delta$ and
$\mu$. Gray line presents the MAE with adaptive parameters. In b) the MAE is presented for the case when the algorithm parameters adaptation is done up to the computer precision.}%
\label{Alg1vsAlg2}%
\end{figure*}

\textit{Example 2}: From the previous example and theoretical considerations we have concluded that adaptive algorithm parameters should be used.
The results obtained by the proposed
method for self-adaptive parameter adjustment are presented in 
Fig.~\ref{Alg1vsAlg2}. The 
criterion (\ref{krit}) for adaptive step size is applied on signal (\ref{SigExample2b}), with 150 randomly positioned
missing signal samples. The starting parameters for the adaptive algorithm were
$\Delta=20$ and $\mu=20$. The graphics in
Fig.~\ref{Alg1vsAlg2}(a) illustrates the MAE as a function of the
iteration number. Each color on these graph matches one set of the parameters
$\Delta$ and $\mu$.
The parameters were divided by 10 (and color in graph is changed) when condition (\ref{krit}) is met.

The dashed color lines represent the MAE when the algorithm with constant parameters is run from the same initial point. 
The solid red line is for the MAE when constant $\Delta=20$ and $\mu=20$ are used. The solid green line is for the MAE when $\Delta=20$ and $\mu=20$ are used at the beginning, while the algorithm has changed the parameters to $\Delta=2$ and $\mu=2$ 
 when condition (\ref{krit}) is met. The dashed green line represents the MAE if $\Delta=2$ and $\mu=2$ are used from the first iteration.
The solid blue line is for the MAE when the algorithm parameters at the beginning were $\Delta=20$ and $\mu=20$, then changed to $\Delta=2$ and $\mu=2$, and finally changed to $\Delta=0.2$ and $\mu=0.2$. The dashed blue line represents the MAE if $\Delta=0.2$ and $\mu=0.2$ were used from the first iteration.
This process continues in the same way two more times for Fig.~\ref{Alg1vsAlg2}(a) and 12 more times for Fig.~\ref{Alg1vsAlg2}(b).

Note that from the behavior of dashed lines (constant parameters) we can conclude that they achieved their stationary state in 10 times more iterations than by using the previous larger parameters. The green dashed line achieved stationary MAE in about 50 iterations, blue in about 500 iterations. So we can expect that the dashed gray line will achieve its stationary MAE in about 5000 iterations, and so on.

We may conclude that the MAE would achieve $10^{-15}$ (what is the standard computer double precision error) in about $10^{14}$ iterations with corresponding constant parameters. Of course this is not acceptable in practical calculations.
As we can see, the same order of MAE is achieved by the presented
adaptive algorithm in  a relatively small number of iterations (about 350).

\begin{figure*}[tp]
\centering
\includegraphics{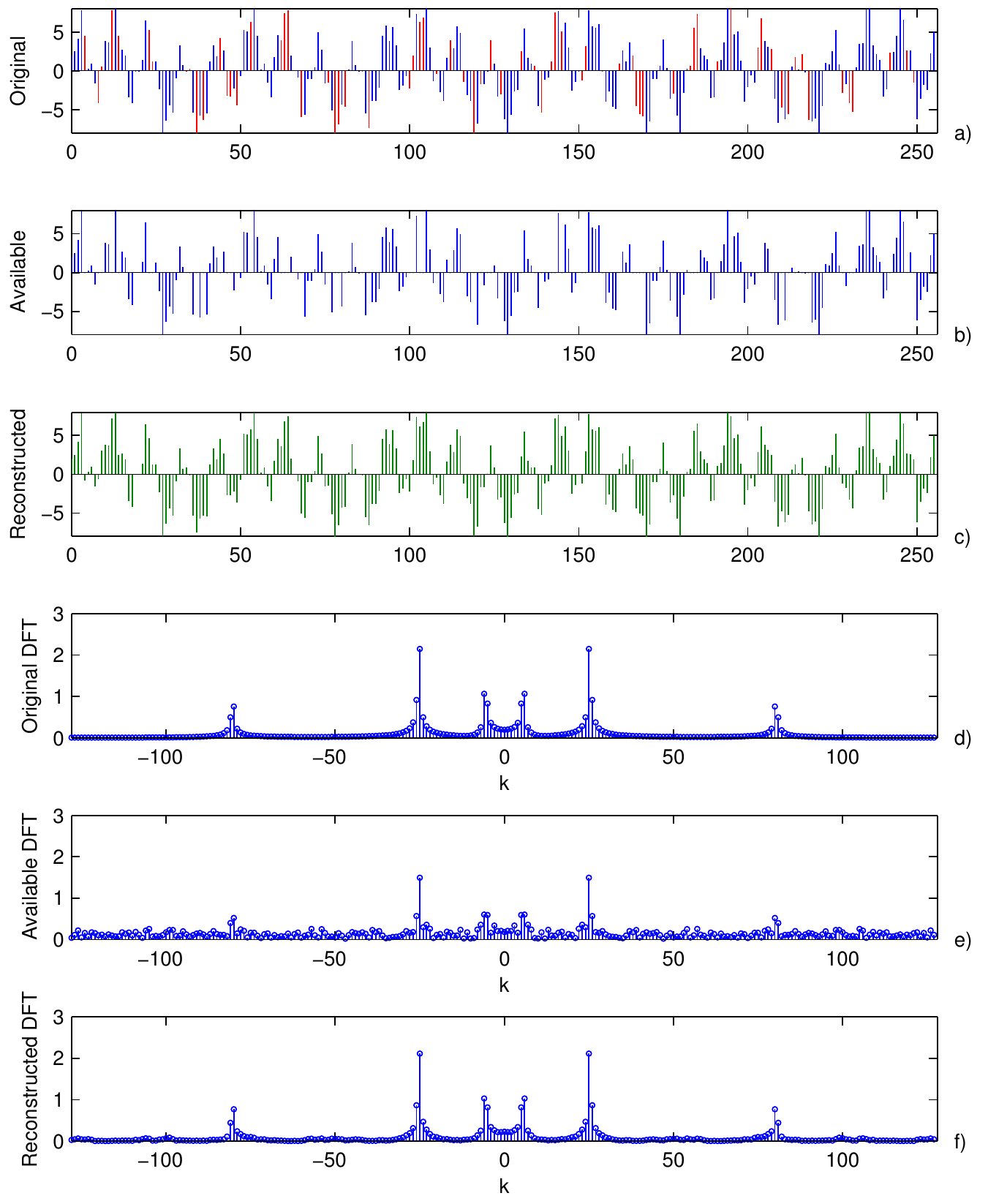}\caption{Reconstruction of an approximately sparse signal whose frequencies do not
match the frequency grid. a) original signal, b) available signal, c)
reconstructed signal, d) DFT of the original signal, e)
DFT of the available signal, f) DFT of the reconstructed signal.}%
\label{grid}
\end{figure*}

\textit{Reconstruction of Approximately Sparse Signals:} Consider a signal:

\begin{align}
x(n)=  &  3\sin(11.2\pi n/N)+5\sin(50.6\pi n/N)\nonumber\\
&  +3\cos(160.8\pi n/N). \label{SigExample}%
\end{align}
whose frequencies do not match frequency grid in the DFT. By definition this
signal is not sparse in the common DFT domain. We will apply the presented
gradient algorithm with $\Delta=3$ and $\mu=4$ on this signal, when 70 signal samples are missing.
Although the analyzed signal is not sparse in a strict sense, satisfactory
reconstruction results are obtained.
Fig.~\ref{grid}(a), (b) and (c) present the original signal, the available
signal samples, and the reconstructed signal, respectively. In 
Fig.~\ref{grid}(d), (e) and (f) the DFT coefficients of the original signal, the
available samples, and the reconstructed signal are shown, respectively. As we
can see, although the original signal is not sparse (its frequencies do not
match to the frequency grid), the reconstruction is good. 

\begin{figure*}[tp]
\centering \includegraphics{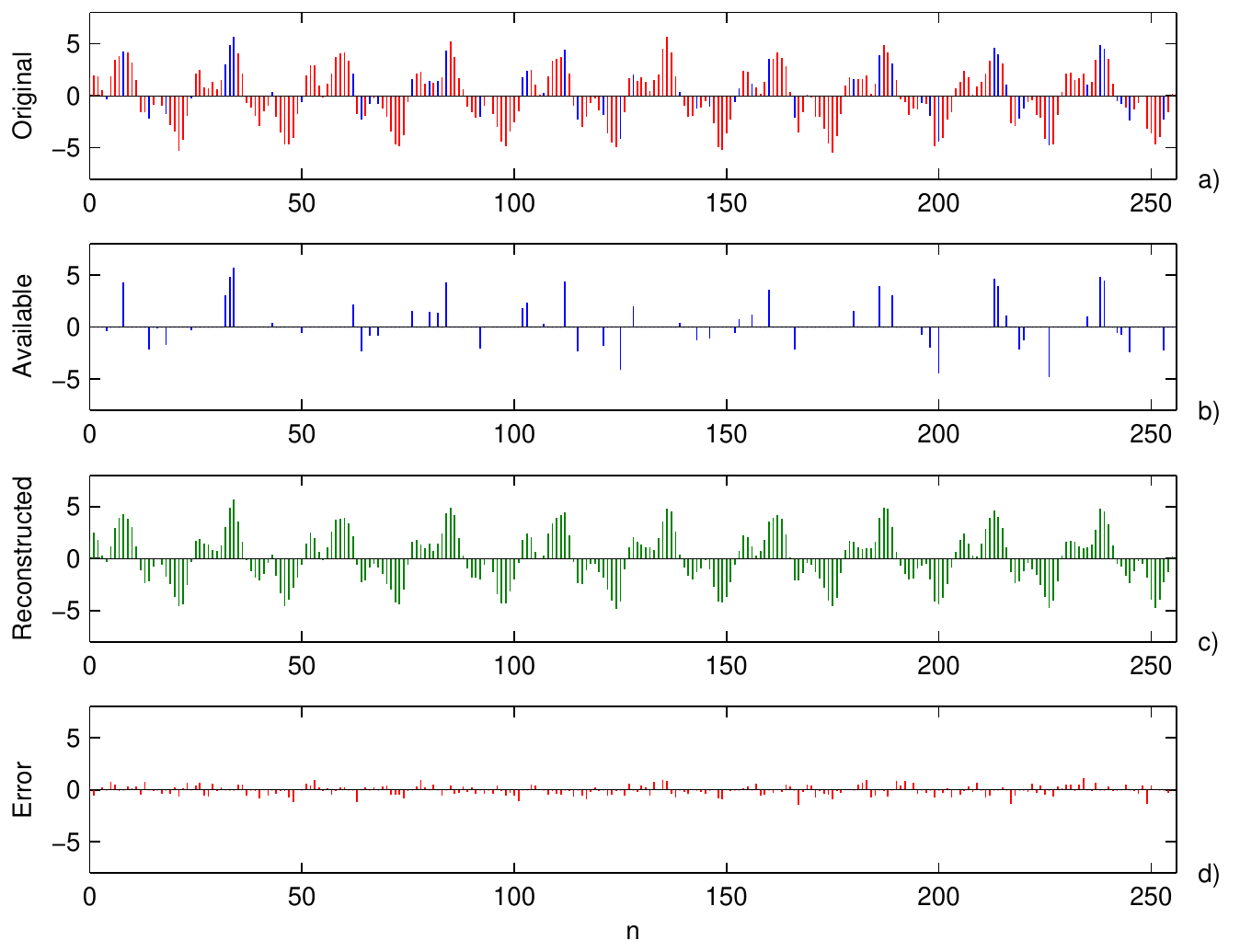} \caption{Reconstruction example for a noisy signal with 200 omitted samples: (a)
original signal; (b) signal with omitted samples set to 0 and used as an input to the
reconstruction algorithm; (c) reconstructed signal; (d) reconstruction error (residual noise).}%
\label{noise200_1}%
\end{figure*}

\textit{Noisy Signals: }The proposed algorithm is used for analysis of a noisy
signal. It has been assumed that a sparse signal is corrupted by an additive
Gaussian noise. From the reconstruction based on a limited number of samples
we can come to a conclusion that in the case of sparse noisy signals certain
improvement can be achieved if a number of signal samples are purposely
omitted and the reconstruction is performed. Reconstruction for 200 omitted samples is presented in Fig.~\ref{noise200_1}.

\begin{figure}[tp]
\centering \includegraphics{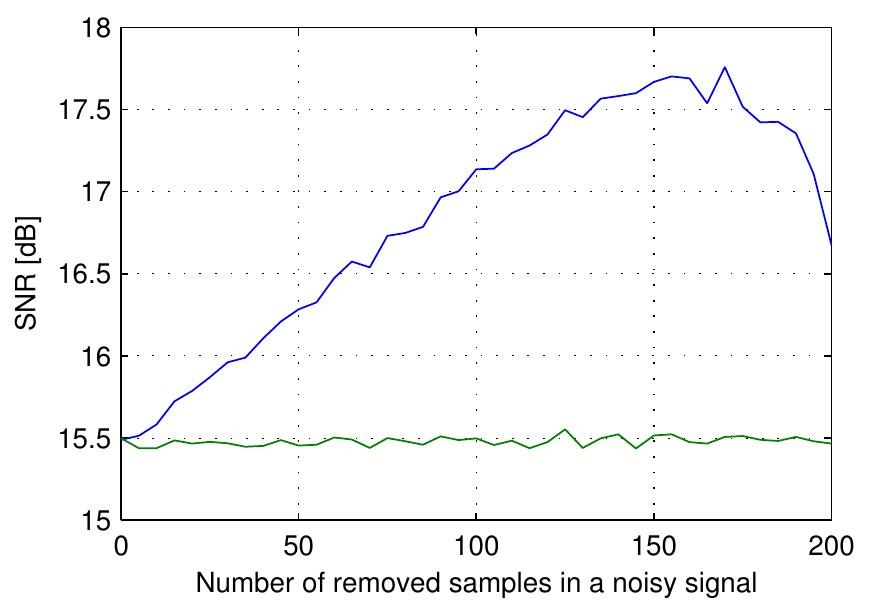} \caption{SNR of the reconstructed signal
as a function of number of omitted samples (blue line). SNR of the original
signal (green line).}%
\label{SNRout}%
\end{figure}

From Fig.~\ref{SNRout} we can see that the SNR improvement higher than $2$dB is achieved if we randomly omit 150 out of 256 signal samples and perform the reconstruction. Improvement could be significantly higher if we were able to selectively remove the most damaged samples \cite{7a}.   

\begin{figure}[tp]
\centering \includegraphics{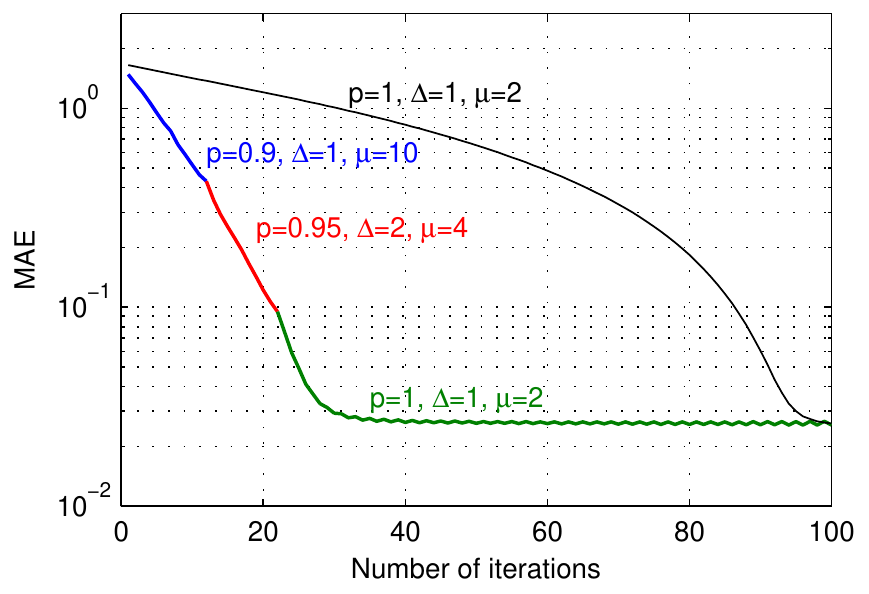} \caption{Reconstruction mean
absolute error for the constant and varying measure and other algorithm parameters.}%
\label{MAE_Delta_Mi_varp_b}%
\end{figure}

\textit{Varying Concentration Measure:} The number of iterations for required
accuracy can be further improved by varying measure parameter $p$. Measures
for $p<1$ are more suitable to gradient based reconstruction. However measures
for $p<1$ do not converge to the true values of missing samples. A possible
solution is to use measures with $p$ slightly lower than $1$ at the beginning of iterative algorithm
and to switch to $p=1$ afterwards. Figure~\ref{MAE_Delta_Mi_varp_b} illustrate
the case when $p=0.9$, $\Delta=1$ and $\mu=10$ is used for iterations 1 to 12,
$p=0.95$, $\Delta=2$ and $\mu=4$ is used for iterations 13 to 22 and finally
$p=1$, $\Delta=1$ and $\mu=2$ is used for iterations form 23 to 100. The case
with constant parameters $p=1$, $\Delta=1$ and $\mu=2$ is presented in the
same figure. 

\section{Conclusion}

In this paper we have presented an algorithm for unavailable/missing samples reconstruction in the sparse signals. The algorithm is based on the concentration measures used to quantify the signal sparsity. Since the commonly used measures are not differentiable around the optimal point, a criterion for variable algorithm parameters is introduced. The presented, gradient-based, adaptive step size algorithm is able to achieve the computer precision accuracy  in a simple and numerically efficient way. The algorithm is applied to sparse signals, including the signals that are only approximately sparse. An example with a noisy signal is considered, demonstrating that some improvements in SNR may be achieved by omitting a number of samples. This algorithm can be applied on any concentration/sparsity measure form. A simple example on this topic is also presented.


\begin{thebibliography}{99}                                                                                               
\bibitem {donoho2006}D.~L. Donoho, \textquotedblleft Compressed sensing,\textquotedblright\ \emph{Information
Theory, IEEE Transactions on}, vol.~52, no.~4, pp. 1289--1306, 2006.

\bibitem {candes2006}E.~J. Cand{\`e}s, J.~Romberg, and T.~Tao, \textquotedblleft Robust
uncertainty principles: Exact signal reconstruction from highly incomplete
frequency information,\textquotedblright\ \emph{Information Theory, IEEE Transactions on},
vol.~52, no.~2, pp. 489--509, 2006.

\bibitem {mallat1999}S.~Mallat, \emph{A wavelet tour of signal processing},
Academic press, 1999.

\bibitem {22}I. Djurovi\'{c}, L. Stankovi\'{c} and  J. F. B\"{o}hme,
\textquotedblleft Robust L-estimation based forms of signal transforms and
time-frequency representations\textquotedblright, \textit{IEEE Trans. on SP},
vol. 51, no. 7, pp. 1753-1761, 2003.

\bibitem {7a}L. Stankovi\'{c}, I. Orovi\'{c}, S. Stankovi\'{c} and M. G. Amin,
\textquotedblleft Robust Time-Frequency Analysis based on the L-estimation and
Compressive Sensing,\textquotedblright\ \textit{IEEE Signal Processing
Letters}, May 2013, pp.499--502.

\bibitem {knjiga}L. Stankovi\'{c}, M. Dakovi\'{c} and T. Thayaparan,
\textit{Time--Frequency Signal Analysis with Application}, Artech House, 2013

\bibitem {Gradient1}M. A. T. Figueiredo, R. D. Nowak and S. J. Wright,
\textquotedblleft Gradient Projection for Sparse Reconstruction: Application
to Compressed Sensing and Other Inverse Problems,\textquotedblright%
\ \textit{IEEE Journal on Selected Topics in Signal Processing}, 2007

\bibitem {Homotopy1}S. Mallat, \textit{A Wavelet Tour of Signal Processing},
Academic Press, San Diego, CA, 1998.

\bibitem {Homotopy2}S. J. Wright. \textquotedblleft Implementing proximal point
methods for linear programming,\textquotedblright\textit{ Journal of
Optimization Theory and Applications}, vol. 65, pp. 531--554, 1990.

\bibitem {Gradient2}T. Serafini, G. Zanghirati and L. Zanni, \textquotedblleft
Gradient projection methods for large quadratic programs and applications in
training support vector machines,\textquotedblright\ \textit{Optimization
Methods and Software}, vol. 20, no. 2--3, pp. 353--378, 2004

\bibitem {L1_magic}E. Candes, J. Romberg and T. Tao. \textquotedblleft Robust
uncertainty principles: Exact signal reconstruction from highly incomplete
frequency information,\textquotedblright\ \textit{IEEE Trans. on Information
Theory}, vol. 52, pp. 489--509, 2006.

\balance

\bibitem {istiterativetresholding}J. More and G. Toraldo, \textquotedblleft On
the solution of large quadratic programming problems with bound
constraints,\textquotedblright\ \textit{SIAM Journal on Optimization}, vol. 1,
pp. 93--113, 1991.

\bibitem {Machinparsuit1}G. Davis, S. Mallat and M. Avellaneda,
\textquotedblleft Greedy adaptive approximation,\textquotedblright%
\ \textit{Journal of Constructive Approximation}, vol. 12, pp. 57--98, 1997.

\bibitem {MP2}D. Donoho, M. Elad, and V. Temlyakov, \textquotedblleft Stable
recovery of sparse overcomplete representations in the presence of
noise,\textquotedblright\ \textit{IEEE Trans. on Information Theory}, vol. 52,
pp. 6--18, 2006.

\bibitem {OMP}B. Turlach, \textquotedblleft On algorithms for solving least
squares problems under an L1 penalty or an L1 constraint,\textquotedblright%
\ \textit{Proc. of the American Statistical Association; Statistical Computing
Section}, pp. 2572--2577, Alexandria, VA, 2005.

\bibitem {18}D. Angelosante, G. B. Giannakis and E. Grossi, \textquotedblleft
Compressed sensign of time-varying signals,\textquotedblright\ \textit{Int.
Conf. on DSP}, 2009, pp.1--8.

\bibitem {5}R. Baraniuk, \textquotedblleft Compressive
sensing,\textquotedblright\ \textit{IEEE Signal Processing Magazine}, vol. 24,
no. 4, 2007, pp. 118--121.

\bibitem {6}P. Flandrin and P. Borgnat, \textquotedblleft Time-Frequency
Energy Distributions Meet Compressed Sensing,\textquotedblright\ \textit{IEEE
Trans. on Signal Processing}, vol. 58, no. 6, 2010, pp. 2974--2982.

\bibitem {16a}L. Stankovi\'{c}, S. Stankovi\'{c} and M. G. Amin, \textquotedblleft Missing Samples Analysis
in Signals for Applications to L-Estimation and Compressive Sensing\textquotedblright,
\textit{Signal Processing}, Elsevier, Volume 94, Jan. 2014, Pages 401--408.

\bibitem {3}F. Ahmad and M.G. Amin, \textquotedblleft Through-the-wall human
motion indication using sparsity-driven change detection,\textquotedblright%
\ \ \textit{IEEE Trans. on Geoscience and Remote Sensing}, vol. 50, no. 12, 2012.

\bibitem {multimedije}S. Stankovi\'{c}, I. Orovi\'{c} and E. Sejdi\'{c},
\textit{Multimedia signals and Systems}, Springer, 2012.

\bibitem {4}Y. Yoon and M. G. Amin, \textquotedblleft Compressed sensing
technique for high-resolution radar imaging,\textquotedblright\ \textit{Proc.
SPIE}, vol. 6968, 2008, pp. 6968A--69681A--10.

\bibitem {20}E. Sejdi\'{c}, A. Cam, L. F. Chaparro, C. M. Steele and T. Chau,
\textquotedblleft Compressive sampling of swallowing accelerometry signals
using TF dictionaries based on modulated discrete prolate spheroidal
sequences,\textquotedblright\ \textit{ EURASIP Journal on Advances in Signal
Processing}, 2012:101 doi:10.1186/1687--6180--2012--101

\bibitem {Ljubisa}L. Stankovi\'{c}, \textquotedblleft A measure of some
time--frequency distributions concentration,\textquotedblright\ \textit{Signal
Processing}, vol. 81, pp. 621--631, 2001

\bibitem {mallat1993}S.~G. Mallat and Z.~Zhang, ``Matching pursuits with
time-frequency dictionaries,'' \textit{Signal Processing, IEEE Transactions on},
vol.~41, no.~12, pp. 3397--3415, 1993.

\bibitem {daubechies2004}I.~Daubechies, M.~Defrise, and C.~De~Mol, \textquotedblleft An
iterative thresholding algorithm for linear inverse problems with a sparsity
constraint,\textquotedblright\ \emph{Communications on pure and applied mathematics}, vol.~57,
no.~11, pp. 1413--1457, 2004.

\bibitem {figueiredo2007}M.~A. Figueiredo, R.~D. Nowak, and S.~J. Wright,
\textquotedblleft Gradient projection for sparse reconstruction: Application to compressed
sensing and other inverse problems,\textquotedblright \emph{Selected Topics in Signal
Processing, IEEE Journal of}, vol.~1, no.~4, pp. 586--597, 2007.

\bibitem {trst2013}L. Stankovi\'{c}, M. Dakovi\'{c} and S. Vujovi\'{c} \textquotedblleft Concentration Measures with an Adaptive Algorithm for Processing Sparse Signals,\textquotedblright\ in \textit{Proceedings of ISPA 2013},
Sept. 4--6, 2013, Trieste, Italy, pp. 418--423

\end{thebibliography}
\end{document}